\documentclass[preprint,pre,aps,superscriptaddress,a4paper]{revtex4}
\usepackage[dvips]{graphicx}
\usepackage[T1]{fontenc}
\usepackage{epsfig,amsmath,amsfonts,amssymb,color}
\usepackage{siunitx}
\usepackage{natbib}
\usepackage{comment}
\usepackage{color}
\usepackage[dvipsnames]{xcolor}
\begin{document}

\title{Molecular dynamics simulation of a binary mixture near the lower critical point}

\author{Faezeh Pousaneh}  
\affiliation{Theoretical Biological Physics, Department of Theoretical Physics, Royal Institute of Technology (KTH), AlbaNova University Center, SE-106 91 Stockholm, Sweden}
\author{Olle  Edholm$^{\scriptscriptstyle{}}$\footnote{Corresponding
    author. Electronic address: \texttt{oed@kth.se}.}  }
\affiliation{Theoretical Biological Physics, Department of Theoretical Physics, Royal Institute of Technology (KTH), AlbaNova University Center, SE-106 91 Stockholm, Sweden}
\author{Anna Macio\l ek}  
\affiliation{Institute of Physical Chemistry, Polish Academy of Sciences, Kasprzaka 44/52, 01-224 Warsaw, Poland}

\begin{abstract}
2,6-lutidine {\color{black}molecules mix} with water at high and low temperatures  but in a wide intermediate temperature range a 2,6-lutidine/water  mixture 
exhibits a miscibility gap.  We constructed and validated an atomistic model for 2,6-lutidine and  performed  molecular dynamics simulations of
 2,6-lutidine/water  mixture at different temperatures.
We  determined the part of  demixing curve with  the lower critical point. The lower critical point
 extracted from our data is located  close to the experimental one. The estimates for critical exponents  obtained from our simulations are 
in a good agreement with  the values corresponding to the  $3D$ Ising universality class.   
\end{abstract}
\maketitle
\section{Introduction}
{\color{black}It is well recognised that } solutions of water with organic molecules may show closed-looped phase diagrams with a miscibility gap. {\color{black} The occurrence of such phase diagrams with more than one critical point 
is very different from what is observed in mixtures of simple fluids. In  simple fluids, the two fluids form a homogeneous mixed  phase  at higher temperatures, whereas they  phase separate  at lower temperatures.
The appearance of an upper critical point (UCP) terminating the two-phase coexistence results  from a competition between the entropy of mixing and  the energy. In the case of mixtures of complex species, such as water and  some organic molecules, the mechanism
behind the occurrence of the miscibility gap is rather involved. As argued in Refs.~\cite{Hirschfelder:1937,Narayanan:1994}},
the existence of a lower critical point (LCP) can be due to formation of directional bonds
between water and  the organic molecules, e.g., hydrogen bonding.
Below the  LCP,   {\color{black}{the}} strong  hydrogen bonding promotes mixing {\color{black} at the expense of rotational degrees of freedom of molecules, which
are effectively "frozen out" by the bonding. At higher temperatures, thermal fluctuations 
``unfreezes'' these rotational degrees of freedom  so that  the hydrogen bonding gets destroyed and   the  mixed phase separates.}
Theoretical studies of simple lattice models and within the Landau-Ginzburg approach have demonstrated that the miscibility gap in liquid mixtures may indeed  emerge due to  angular dependent 
attractive interactions on top of the spherically symmetric ones~\cite{Walker:1980,Walker:1987,Walker:1983,Narayanan:1994,Almarza:2012}. 
Experiments show that the shape of  closed-loop phase diagrams of aqueous solutions of organic molecules near the LCP 
is very flat, i.e., concentrations of the species in the two coexisting phases near LCP vary strongly with temperature. 
Such behavior indicates   that in both coexisting liquid phases some local structures are formed which are determined by hydrogen bonding between water and the organic molecules. 
{\color{black}The features of the miscibility gap}, in particular its sensitivity  to changes  {\color{black}of} the intermolecular interactions,  has been studied
{\color{black}theoretically~\cite{Brovchenko:1997}, by computer simulation (for tetrahydrofuran-water mixtures) ~\cite{Brovchenko:2001,MDS:2016}, and experimentally~\cite{Brovchenko:1997,Robertson:2015} } by several techniques, e.g.  by deuteration of water,  addition of electrolytic impurities or hydrotrops.
{\color{black}As follows from these  studies}, the  details of the interactions {\color{black}affect} the critical temperature as well as  the detailed shape of the phase diagram. 

The aqueous solution of 2,6-lutidine  (2,6-dimethylpyridine) is a  binary liquid mixture, which has gained considerable attention
in context of its wetting behaviour at silica walls ~\cite{Pohl_et}, porous glass~\cite{Frisken:1991}, and colloids ~\cite{Broide:1993,Gallagher:1992}.
Of particular interest has been the effect of temperature changes on the reversible
aggregation of colloidal particles dispersed in a 2,6-lutidine/water mixture~\cite{Beysens:1985,Kurnaz:1995,Kurnaz:1997}. 
More recently, 2,6-lutidine/water mixtures were used  to determine 
critical Casimir interactions in colloidal systems 
\cite{Hertlein-et:2008,Gambassi-et:2009,Nellen-et:2009,Nguyen-et:2013}. 
The reason for the popularity of 2,6-lutidine/water mixture  in  these studies is that it possesses a closed loop phase diagram with a relatively wide temperature miscibility gap, i.e.,
the difference between the upper and lower critical point is large ($\approx$197$^\circ\textrm{C}$). Further, the LCP is  conveniently located  near room temperature.
The   phase diagram as well as static and dynamic critical properties of 2,6-lutidine/water were studied  
intensively  experimentally~\cite{andon:52:00,stephenson:93:00,Grattoni:93:00,cox:56:00,Jayalakshmi:1993}.
The LCP has been reported to  occur {\color{black}at }
$T_c\approx 307.1~$K and at the lutidine  mole fraction $x_{lut}\approx 6.1\% $ ~\cite{andon:52:00,Grattoni:93:00,cox:56:00,francis:61:00}.
The location of the LCP is, however,
strongly affected by impurities and, moreover,  it is  sensitive to the method of determination. 
 This is why the values of  the lower critical temperature and the critical mass, volume or mole fraction
  vary in the literature~\cite{Loven:1963,Guelari:1972,Levchenko:1993,Jayalakshmi:1993,Mirzaev:2006}. Such uncertainty is especially troublesome for application of a 2,6-lutidine/water liquid mixture as  solvent in colloidal systems tuned by critical Casimir interactions, where the precise knowledge of the deviation 
  in temperature and concentration from the critical values is required. Computer simulations of the 2,6-lutidine/water mixture are thus highly desirable. {\color{black} Moreover, such simulations
  can help to better understand the molecular mechanisms behind the lower critical point,} and {\color{black}are} a necessary prerequisite for studies {\color{black} of} more  complicated phenomena such as, for example, formation of  of mesostructures 
in a mixture of water/organic solvent by adding an antagonistic salt, which is composed of hydrophilic cation and hydrophobic
anion.
Such mesostructures were observed recently in SANS experiments {\color{black}in} a mixture of water/3-methylpyridine/NaBPh$_4$ near the lower critical point \cite{sadakane:11:00} and off-critical point \cite{sadakane:09:00,leys:13:00}.
A similar observation is reported for  2,6-lutidine/water mixture 
 \cite{sadakane:14:00}. These phenomena are only partially explained {\color{black}theoretically}  \cite{pousaneh:14:00,Onuki:04:00}.
Another interesting topic, which can  be an extension of the present work  is  critical adsorption of  2,6-lutidine/water mixture containing salt (inorganic one)  at a  charged and selective  wall. This phenomenon is of crucial relevance to recent experiments on critical Casimir interactions \cite{Hertlein-et:2008,Nellen-et:2009,nellen:11:02}. The findings of these experiments were not yet clarified in a satisfactory manner and there  is {\color{black}some}  controversy about {\color{black}their}  origin. Some analytical {\color{black} studies of} this problem are available in the literature \cite{ciach:10:00, pousaneh:12:00,bier:11:02}. However, these results were obtained within {\color{black} a} mesoscopic model  and by using  various approximations and thus they need to be verified by means of    microscopic simulations.

     In the present paper we propose an atomistic description of the 2,6-lutidine molecule and apply it to study the bulk 2,6-lutidine liquid as well as the 2,6-lutidine/water mixture near the LCP by molecular dynamics simulations. 
 The goal is to check whether our   model of the 2,6-lutidine molecule is able to capture the main features
  of the  bulk fluid as well as those of the aqueous  solution.

\section{Computational details, modelling and validation of the model}

\subsection{Computational details}
\label{computationaldetails}
All  simulations were performed by the Gromacs/4.6.7 
 package~\cite{spoel:05:00}. The Gromos54a7 force field~\cite{Schmid:11:00}  was applied for Lennard-Jones (LJ)  pair potential parameters, bond lengths and bonded parameters   for angles, and dihedrals. The Particle Mesh Ewald (PME) approach~\cite{Darden:93:00} was applied for electrostatic interactions, while a cut-off length  $r_c=1.2$ nm was applied to the LJ interactions.  Simulations in this work were either  performed  in the NpT ensemble (constant pressure) or NVT ensemble (constant volume); the type of  ensemble  is mentioned in the text in  each case. 
 In the NpT ensemble, the temperature and the pressure  were controlled by a V-rescale thermostat  \cite{Bussi:07:00}   and a Parrinello-Rahman barostat  \cite{Parrinello:81:00}  (isotropically to $p=1$ atm), respectively. For the NVT ensemble, the temperature was  controlled by V-rescale and no pressure coupling was applied.  All bond lengths were  constrained with the LINCS algorithm \cite{Hess:97:00}. For  water the TIP4P/2005 model was  used \cite{Abascal:05:00}. {\color{black} The simulation outcomes were analyzed through our own programs, Gromacs  and VMD plugins  \cite{VMD,Giorgino}.}
\begin{figure}
\centering
\includegraphics[scale=0.3]{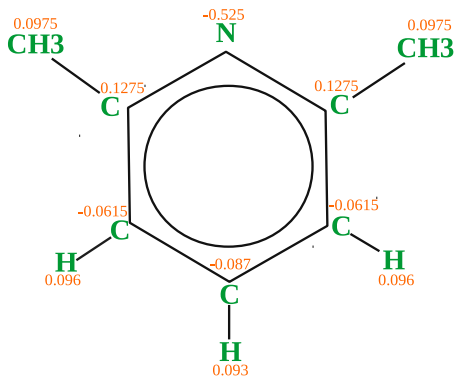}
\caption{The 2,6-lutidine molecule  with charges from the final parametrisation used in the current work. The molecule is modelled by 11 atoms where CH$_3$-groups are considered as united atoms.}
\label{model}
\end{figure}
\subsection {Parametrisation of the 2,6-lutidine molecule}
We represent the  2,6-lutidine  molecule, C$_7$H$_9$N, by 11 atoms wit the  two  CH$_3$ groups treated as  single united atoms   as shown in Fig.~\ref{model}, while the  hydrogen atoms that are attached to the ring carbons are explicitly included.
{\color{black}
The GROMOS force field does not provide partial
charges on 2,6-lutidine molecule, therefore, we obtained an initial estimation of these
charges from quantum chemistry simulations. Then, we  varied the values of these charges
until two goals were achieved: (i) an agreement with the experimental results for the
density $\rho$ and the heat of vaporization $ \Delta H_{vap}$ for liquid lutidine at temperatures around
room temperature, and (ii) the existence of the LCP for the mixture of 2,6-lutidine and water
in the experimental range (details of 2,6-lutidine/water simulation will be given in Subsec.~\ref{w+l}). More precisely, we gradually scaled the  partial charges of 2,6-lutidine molecule by a factor in
order to change its Coulomb interaction with water until we  obtained the LCP temperature close
to experimental value. From experiments, we know that  we should have  a mixture at $T = 280$ K
and a  phase-separated system at $T = 320$K. Therefore, for each rescaled charge distribution,  simulations were carried out at these two
temperatures. 
The appearance of  mixed and phase-separated phases  at these two temperatures ensures  that the LCP is located somewhere in between.
During the rescaling, we  also took advantage of  the fact that the molecule is symmetric and that the
total charge is zero. Moreover, we  kept the charges on the hydrogen atoms
and  CH$_3$-groups consistent with the Gromos force field and just vary the remaining 3
parameters.}

\subsection {Validation of the lutidine model}
Simulations were performed
 for $80~$ns using a $2~$\SI{}{\femto s}  time step in the NpT ensemble for the bulk system and  in the NVT  ensemble for the surface tension calculation. Results collected  after an equilibration of 10 ns, are presented   together with the  corresponding   experimental data taken from Refs.~\citep{Steele:86:0,Jenkins:90:00, bakshi:93:00,NIST,Althof:97:00} in Table.~\ref{tab1}. 
The simulations provide data for the  density $ \rho$, the static dielectric constant $\epsilon$ and the heat of vaporization  $\Delta H_{vap}$  in a  fair agreement with the experimental ones.
The heat of vaporization was calculated as
~\cite{{feig:10:00}}
\begin{eqnarray}
\Delta H_{vap} =  \langle U_{gas}\rangle - \langle U_{liquid}\rangle +RT,
\end{eqnarray}
where $\langle \cdot \rangle$ denotes time average.  $U_{gas}$ and $U_{liquid}$ are potential energies of lutidine in the gas and liquid phases. The gas phase is considered as ideal so $U_{gas}$ 
contains  {\color{black} only  interactions within the molecules}. The surface tension $\gamma$ was obtained from simulations of the liquid with two flat surfaces. This  system  was created by increasing the periodic box size of the equilibrated bulk system by one order of magnitude in one direction. The usage of a 1.2~nm cut-off for the LJ interactions has resulted in a too small surface tension.   Therefore, we performed 
 simulations of this system using   PME treatment of  LJ interactions (LJ-PME). This  gave the   surface tension of $32.5 \pm 0.2~$ mN/m in a better agreement with 
experiment. With the  fractional charges used in the  present model, the dipole moment of lutidine is $2.5~$D. The experimental gas phase value is $1.7~$D (no experimental value is available for  liquid).  Similar differences are encountered  for most water models;  typically they predict about 30\% larger dipole moment than possessed by water in a gas phase, although the difference here is slightly bigger. These differences are caused   by the mutual polarization of molecules in the liquid  phase. The good agreement between the experimental dielectric constant and the one calculated from fluctuations in the total dipole moment of the entire simulation box is reassuring and indicates that the electrostatic properties of the 2,6 lutidine fluids are properly modelled.  

Finally, the heat capacities at constant pressure and constant volume, $C_p$ and $C_V$,  were calculated from the enthalpy and energy fluctuations in the simulations as
\begin{eqnarray}
C_p=\frac{\sigma^2_H}{k_BT^2}\;\;\;\;\mbox{and}\;\;\;\;C_V=\frac{\sigma^2_E}{k_BT^2}.
\end{eqnarray}
The calculated value for $C_p$ given in the last row of the  Table.~\ref{tab1} shows  a substantial difference  compared to the experimental value. The reason for this is that the {\it{classical}} treatment 
of the lutidine molecules  allows too much energy to be taken up by degrees of freedom that in reality are  {\it quantum mechanical}.
A quantum-correction could, however, be added to the classical heat capacity assuming that the relevant degrees of freedom could be approximated as coupled harmonic oscillators.
To do this we follow Refs.~\cite{berens:83:00}-\cite{waheed:11:0} and  determine the normal modes of the system from the velocity auto-correlation  functions. 

\begin{table}
\centering
\begin{tabular}{|c|c|c|c|}
\hline
     Quantity& T[K]  & Simulation & \hspace{0.5cm} Experiment      \\
\hline
  $\rho$ [kg/m$^3$] & 298   &    $940\pm0.4$    &  925   \\
\hline
$\epsilon$ & 298  &  $7.1\pm0.2$   &   6.9  \\  
  \hline
$\Delta H_{vap}$ [kJ/mol]  & 298  &  $45.4\pm0.2$   & 43.7$-$46.1$ $\\
   \hline
$\gamma$ [mN/m],\hskip0.5cm $r_c^{LJ}=1.2$ nm & 307  &   $21.1\pm0.4$ & 29.8   \\
 \hskip1.5cm  LJ-PME  &307 & $32.5\pm0.2\;$  &     \\
\hline
  $C_p$ [J mol$^{-1}$ K$^{-1}$] & 298  &   $260\pm3$   &   183   \\
  \hline
\end{tabular}
\caption{Simulation results and available experimental values for various physical quantities characterizing  2,6-lutidine.}
\hspace{1cm}
\label{tab1}
\end{table}

The normal mode distribution $S(\nu)$ was obtained from the Fourier transform of the velocity correlation functions and the quantum correction to the heat capacity 
$C_V^{QM.corr}$ was obtained as the difference between the heat capacity of a quantum oscillator and a classical oscillator ($k_B$) integrated over the normal mode distribution
\cite{berens:83:00}
\begin{eqnarray}
\label{QM}
C_{V}^{QM.corr}=k_B \int _0^\infty d\nu S(\nu) W_{c_V}(\nu);\hspace{1cm}
 W_{c_V}(\nu)=\Bigg( \frac{u^2 e^u}{(1-e^u)^2}-1 \Bigg),
\end{eqnarray}
with  $u\equiv h \nu/k_BT$ being the energy in thermal units. \\
The quantum corrections to the heat capacities at three temperatures were obtained according to  Eq.~\ref{QM} by using the normal mode distributions obtained  from simulation of $N=1104$ molecules of 2,6-lutidine at the different temperatures. The simulations were done in the NVT ensemble for $5~$ns and velocities were stored  every $5~$fs. 
Although the quantum corrections were calculated at constant volume, they can still be applied to the  constant pressure heat capacities, {\color{black}assuming} that the difference between quantum corrections in the two ensembles is negligible.   
The  classical heat capacities $C_{V,p}^{class}$  was calculated from the fluctuations in energy and enthalpy calculated from the last $12~$ns of $20~$ns-long simulations at constant volume or constant pressure, respectively. There are a few additional contributions to the heat capacity  due to quantum mechanical vibrations in the bond lengths which were treated rigidly in the simulations, and due to the quantum mechanical motion of the absent hydrogen atoms in the CH$_3$ groups. These contributions are negligible.  
The corrected heat capacity $C_{V,p}^{corr}$ was obtained as the sum of the classical value and the quantum correction 
\begin{eqnarray}
C_{V,p}^{corr}=C_V^{QM.corr}+C_{V,p}^{class}.
\end{eqnarray}
In the last two columns of Table.~\ref{tab2}, the obtained quantum corrected heat capacities $C^{corr}_p$ and the experimental ones can be compared. 

\begin{table}
\vspace{1cm}
\begin{tabular}{|l|  c c c|c  c|   c|   }
\hline
   & NVT &  & &   NpT    &&    \\
   \hline
$T$ [K]  &$C^{class}_V$ & $C^{QM. corr}_V$ &$C^{corr}_V$ & $C^{class}_p$ &$C^{corr}_p$&$C^{exp}_p$  \\ 
\hline
280 & $191\pm 2$ &$-67$ & $124\pm 2$& $256\pm2$ &  $190\pm2$  &-\\
\hline
300 & $188 \pm 2$ & $-63$ & $125\pm 2$ & $260\pm2$ & $197\pm2$&  184\\
\hline
333 &$185\pm 1$ &$-57$ & $128 \pm 1$ & $266\pm 3$ &  $209 \pm3$  &   196\\
\hline
\end{tabular}
\caption{Simulation data for the classical and the quantum corrected   heat capacities in J mol$^{-1}$ K$^{-1}$. The corrected heat capacities $C^{corr}_V$ and $C^{corr}_p$ are given and  $C^{corr}_p$ can be compared to the available experimental values given in the last column.}
\label{tab2}  
\end{table}

For most fluids (as well as other condensed matter systems) $(C_p-C_V)/C_V$ is much smaller than 1. For liquid water for instance, this quotient is about 0.01. It is therefore a bit surprising  that this quotient is as large as about 0.5 for the present system. As a consistency check, we  use the exact thermodynamic relation 
\begin{eqnarray}
C_{p}-C_V= V T \frac{\alpha_p^2}{\kappa_T}
\label{cpcveq}
\end{eqnarray}
to calculate  $C_p-C_V $ for lutidine from the molar volume,  coefficient of thermal expansion $\alpha_p= \frac{1}{V} \frac{\partial V}{\partial T }|_p$ and the isothermal compressibility  $\kappa_T=-\frac{1}{V} \frac{\partial V}{\partial p }|_T$. The thermal expansion coefficient and the isothermal compressibility were obtained from the fluctuations in the NpT simulations. The isothermal compressibility was obtained from the volume fluctuations, while the thermal expansion coefficient was obtained from the cross correlations between volume and enthalpy fluctuations. The appropriate equations are discussed  in Ref.~\cite{Allen:87:00}  
  and are 
\begin{eqnarray}
\kappa_T=\frac{1}{Vk_BT} \langle(V-\langle V\rangle)^2\rangle \;\;\;\mbox{and}\;\;\;\alpha_p=\frac{1}{Vk_BT^2} \langle(V-\langle V\rangle)(H-\langle H\rangle)\rangle. 
\end{eqnarray}
 The data from the simulations shown in Table.~\ref{cpcv} are consistent with the data in Table.~\ref{tab2}. For comparison, the experimental data for liquid water (taken from ~\cite{Haynes}) are  also shown in the table.  The table shows  that the main reason for the  big difference 
between the two  heat capacities for lutidine  is the large thermal expansion coefficient (which is squared in the Eq.~\ref{cpcveq}).  
It is worth mentioning that the similar big difference between $C_p$  and $C_V$ occurs in liquid benzene ~\cite{mortimer:08:00} and is reported experimentally in Ref.~\cite{smith:65:00}. Experimental data for benzene are also given in the table. 

\begin{table}
\vspace{1cm}
\scalebox{0.9}{
\begin{tabular}{|l|l|  c   |c| c| c| c| c   }
\hline
Substance &        $T$ [K]  & $\alpha_p$ [K$^{-1}$]  & \hspace{0.3cm}$\kappa_T $ [Pa$^{-1}$] & \hspace{0.3cm} $V$[m$^3$/mol] &  \hspace{0.3cm} $ V T \frac{\alpha_p^2}{\kappa_T}$[J mol$^{-1}$ K$^{-1}$]  & \hspace{0.3cm} $(C_p-C_V)$[J mol$^{-1}$K$^{-1}$]  \\
\hline
2,6-lutidine &280 & $1.67\cdot 10 ^{-3}$   &   \hspace{0.3cm} $1.38\cdot 10 ^{-9}$  &  \hspace{0.3cm} $1.11\cdot 10 ^{-4}$  &  \hspace{0.3cm} 62.8 &65\\
\hline
2,6-lutidine &300 & $1.75\cdot 10 ^{-3}$   &   \hspace{0.3cm} $1.56\cdot 10 ^{-9}$  &  \hspace{0.3cm} $1.14\cdot 10 ^{-4}$  &  \hspace{0.3cm} 67.1  &72\\
\hline
2,6-lutidine &333 & $1.93\cdot 10 ^{-3}$   &   \hspace{0.3cm} $1.85\cdot 10 ^{-9}$  &  \hspace{0.3cm} $1.18\cdot 10 ^{-4}$  &  \hspace{0.3cm} 79.1 &81\\
\hline
water &300 & $0.274\cdot 10 ^{-3}$   &   \hspace{0.3cm} $0.45\cdot 10 ^{-9}$  &  \hspace{0.3cm} $0.18\cdot 10 ^{-4}$  &  \hspace{0.3cm} 0.9 &0.9\\
\hline
benzene &293 & $1.23\cdot 10 ^{-3}$   &   \hspace{0.3cm} $0.967\cdot 10 ^{-9}$  &  \hspace{0.3cm} $0.891\cdot 10 ^{-4}$  &  \hspace{0.3cm} 41.3 & 40.5 \\
\hline 
\end{tabular}}
\caption{ A  comparison between $ V T \frac{\alpha_p^2}{\kappa_T}$  and $C_p-C_V$ in J mol$^{-1}$ K$^{-1}$, for lutidine (simulations), water and benzene (experiments).}
\label{cpcv}  
\end{table}

\begin{figure}[t]
\center
\includegraphics[scale=0.9]{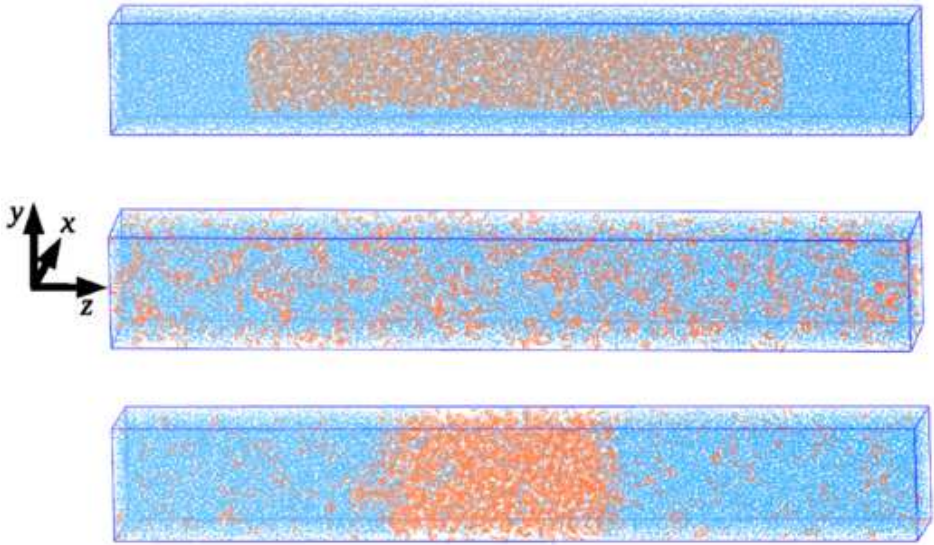}
\vspace{-0.7cm}
\caption{ Snapshots of the initial configuration of the  2,6-lutidine/water  simulations   (top), simulation result at $T=280$ K (center) and  $T=380~$K (bottom). The orange and blue colors indicate the 2,6-lutidine and  water molecules, respectively.}
\label{snapshots}
\end{figure}

\subsection{Simulations of the 2,6-lutidine/water  mixture}
\label{w+l}
In this section we present the simulation results for  the  2,6-lutidine/water  mixture.
{\color{black} 
In order to simulate a 2,6-lutidine/water  mixture, we took  as  initial configuration a box 
of the size ($L, L, 7L$), with $L\approx 3.8 ~$nm,  containing $N_{lutidine}=2050$ equilibrated bulk lutidine molecules, 
placed it at the center of the periodic  box  ($L, L, 7L$) with $L\approx 5.8 ~$nm and  filled (solvated)  with  $N_{water}=31325$ equilibrated water molecules   described by TIP4P/2005 model.  This configuration  corresponds to a lutidine  mole fraction  $x_{lut}=  6.14 \% $, which is close to the experimental value at the LCP. 
This initial configuration, which is neither a mixed phase nor a two-phases mixture, is shown in the top panel of Fig.~\ref{snapshots}; this configuration we  used  for  all studied temperatures here. 
The non-cubic shape of  the box with one side much longer than the other two sides has been chosen  for two reasons. Firstly, we want   ''planar interfaces'' separating  lutidine-rich and lutidine-poor phases for  temperatures as close to the critical temperature $T_c$ as possible.
As discussed in details in Refs.~\cite{godavat,shi:2011,MacDowell:2006}, the larger the  size ratio of the rectangular box, the closer one can approach $T_c$  keeping the   slab structure of the lutidine-rich phase  and, hence, the planar interface.
\begin{figure}
\centering
\includegraphics[scale=0.6]{G,380.eps}
\includegraphics[scale=0.16]{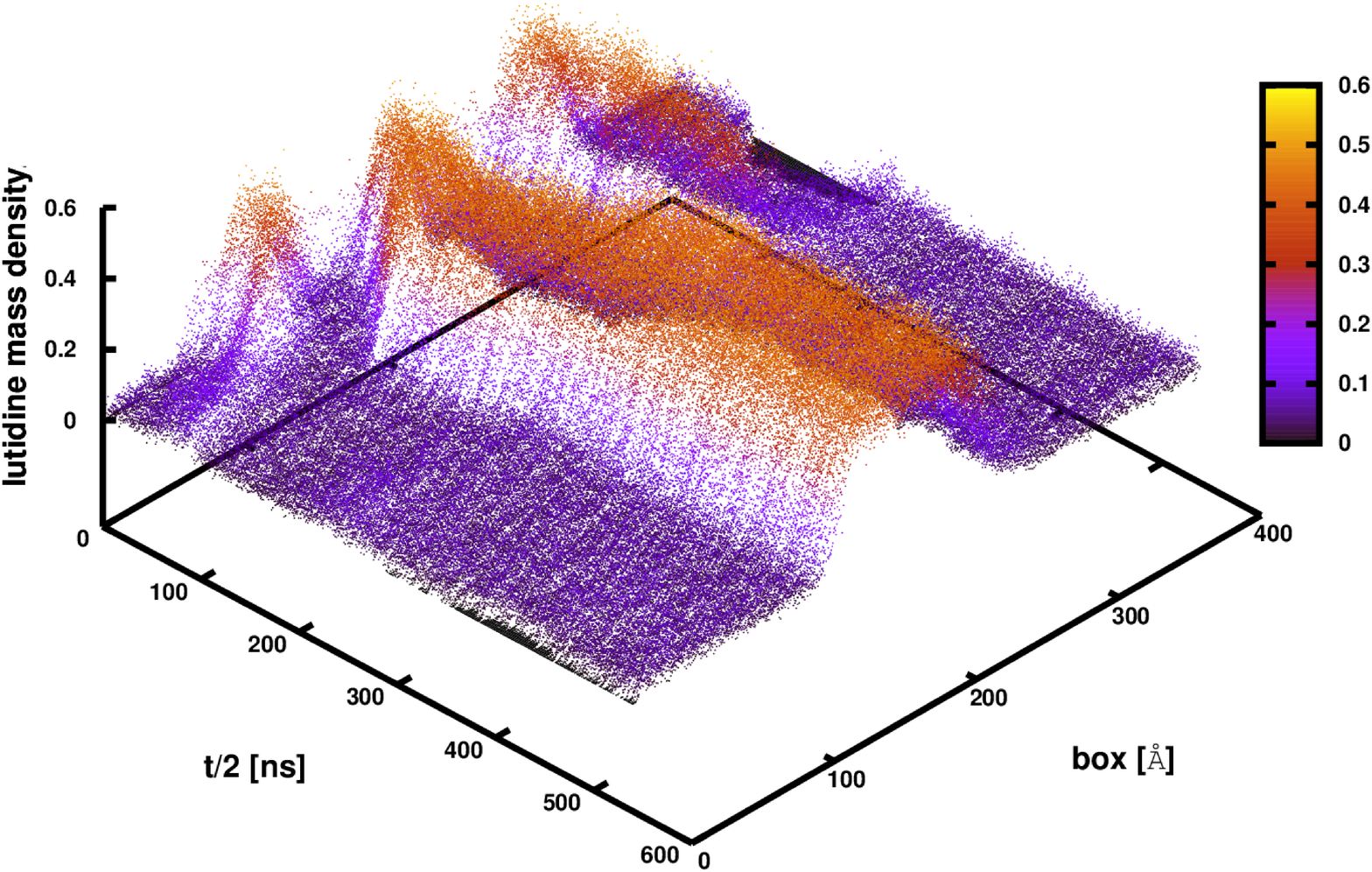}
\vspace{1cm}
\hspace{10cm}
\includegraphics[scale=0.6]{G,320.eps}
\includegraphics[scale=0.16]{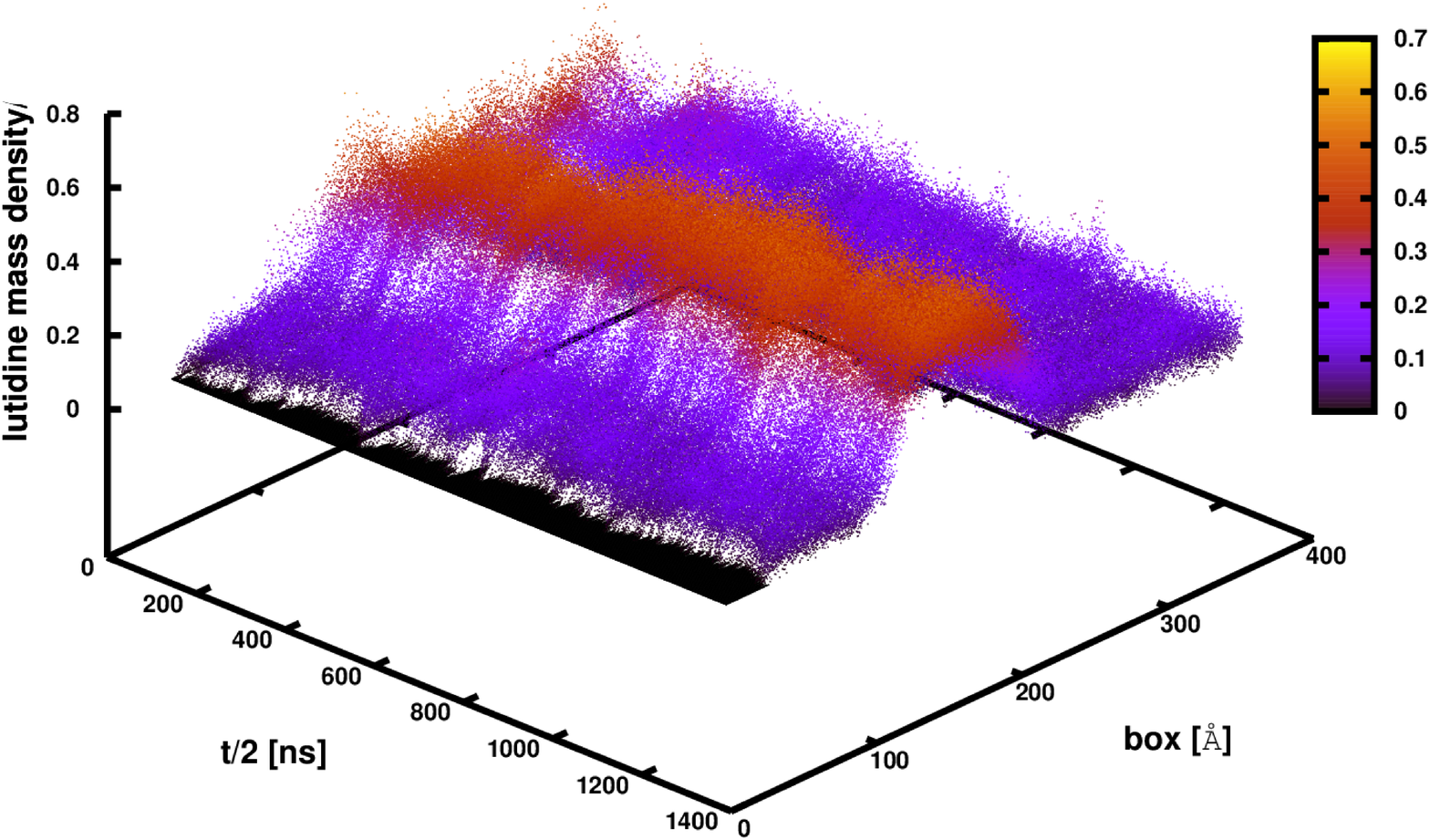}
\vspace{1cm}
\hspace{10cm}
\includegraphics[scale=0.6]{G,280.eps}
\includegraphics[scale=0.16]{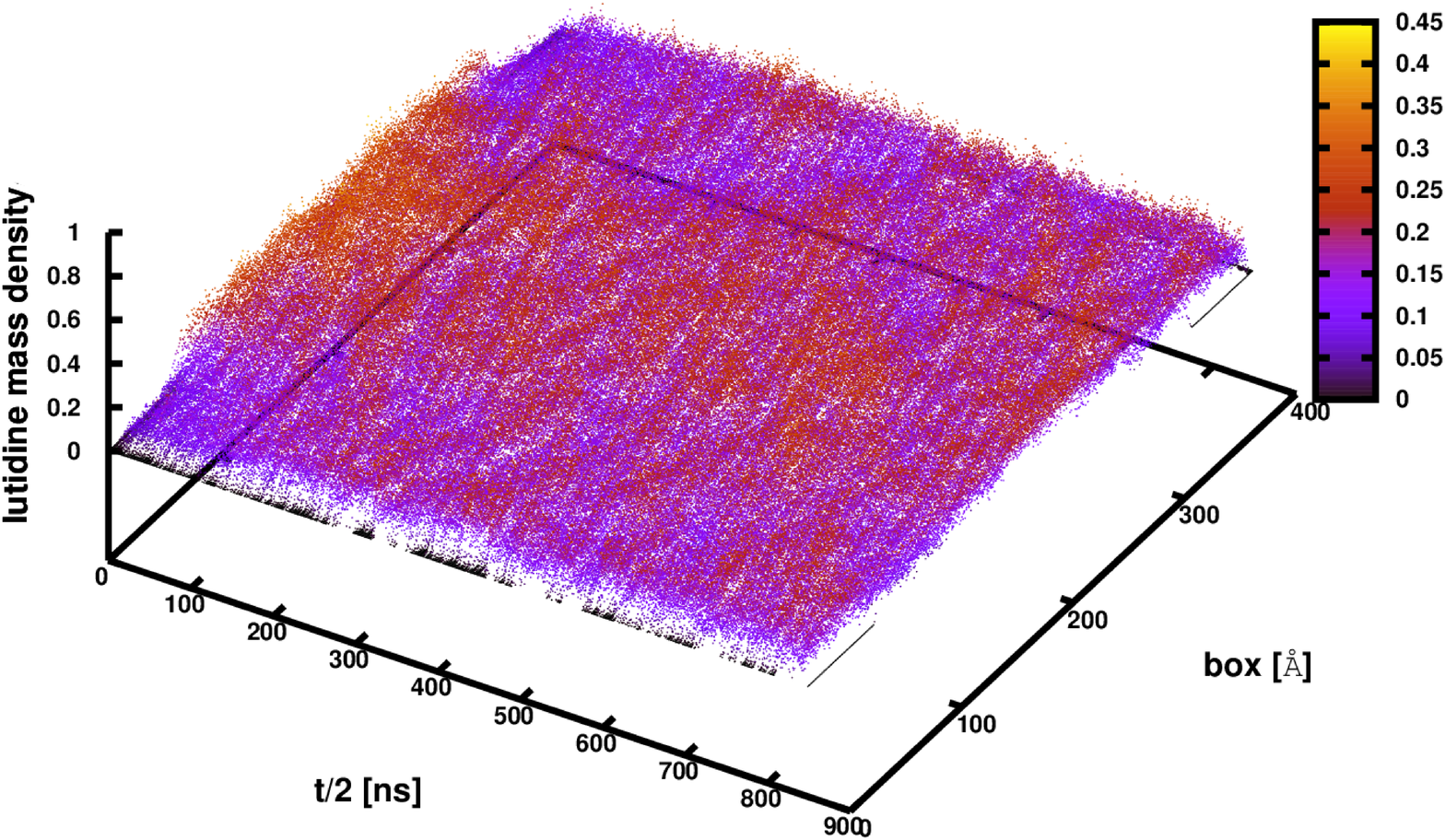}
\caption{(top) left: Time evolution of the lutidine mass density at two different coordinates  $z$ of the simulation box  corresponding to two phases, right: Time evolution of the lutidine mass density versus $z$ coordinate of the box  at $T=380 K$ . (center) the same as in the top panel but  at   $T=320 K$. (bottom) left:  Time evolution of the lutidine mass density at several different coordinates  $z$, right: Time evolution of the lutidine mass density versus $z$ coordinate of the box  at  $T=280 K$. Mass densities are in unit Da/{\AA}$^3$.   }
\label{evol}
\vspace{1cm}
\end{figure}
If the size of the box in the $z$ direction is not big enough,  the slab structure is not stable close to $T_c$. Rather, the lutidine-rich phase  forms a cylinder or a sphere. Within the  simulation  box that we have chosen, all volume fractions at all considered temperatures produce a  planar interface.
Secondly, we want the finite-size effects to be small in order to be able to determine the near-critical properties of a system as accurately as possible~\cite{chepela:00:00,holcomb:00:00,chen:00:00,trokhymchuk:00:00}. This is achieved by choosing the size of the box  to be larger than the bulk correlation length  for all studied temperatures.
Moreover, due to the enlargement of the periodic  box in the $z$-direction the  two  interfaces in the slab structure do not influence each other.
}

\begin{figure}
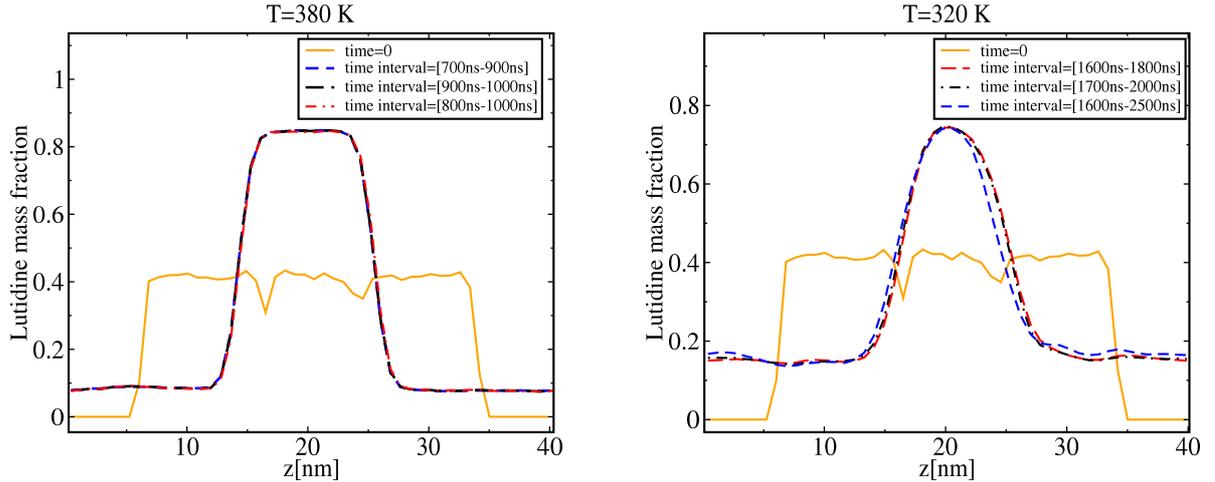

\centering
\includegraphics[scale=0.7]{ref_380.eps}
\hspace{0.8cm}
\includegraphics[scale=0.7]{ref_320.eps}
\caption{{\color{black} The lutidine mass fraction versus the $z$ coordinate of the simulation box for temperatures $T= 380$ and $ 320$~K. The plots show the initial configuration and three time intervals after the equilibration.} }
\label{density_1}
\end{figure}

We simulated the 2,6-lutidine/water mixture in the NpT ensemble for various temperatures;  the simulation setting is given  in Sec.~\ref{computationaldetails}. 
{\color{black}
The time evolution of the local mass density of 2,6-lutidine for several temperatures  are presented in Fig.~\ref{evol}. The plots show clearly a  phase separated system  at $T=320$ and  $380~ K$ and a homogeneous mixture at $T=280 K$. 
 The actual equilibration time is determined with the time at which  not only   the density profiles  remain the same within the statistical errors, but also the energies and all hydrogen bonds become stable.
 Fig.~\ref{density_1} presents the  density profiles averaged  over  different time intervals  after the equilibration, for two temperatures. Depending on temperature, stable equilibrium structures were reached after $0.7-3.7~\mu s$. The simulations took several months, with average run of $50 ~(ns/day)$ for each temperature.
}

The results of the simulation are reported in the next section. 

\section{Simulation results}

\subsection{Phase behaviours of the  2,6-lutidine/water  mixture near the lower critical point}
Fig.~\ref{snapshots} shows the snapshots of the initial configuration for all simulations (top)  and simulation results for temperatures $T=280$K (center) and $T=380~$K (bottom). 
The snapshots show that the two fluids mix at $T=280~$K, while they phase separate at the higher temperature $T=380~$K.
{\color{black} 
In order to assure that the phase separation is not effected by the initial configuration, the simulations at these two temperatures were redone with different initial configurations. Although we used a mixed phase as the  initial configuration for the higher temperature  $T=380~$K and a phase-separated initial configuration for the lower temperature $T=280~$K,  the same final configurations as  given in   Fig.~\ref{snapshots} (bottom and center respectively) were reproduced.
}

\begin{figure}
\vspace{1cm}
\center
\includegraphics[scale=0.7]{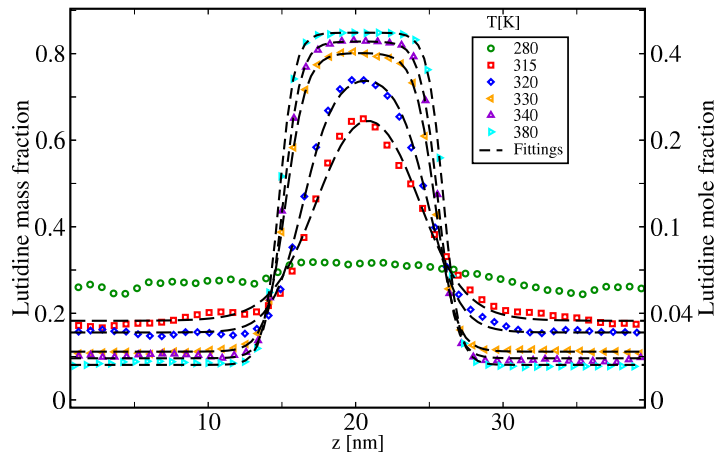}
\caption{Mass fraction (left axis) and mole fraction (right axis) of 2,6-lutidine obtained from  simulations (symbols) fitted to the analytical expression Eq.~\ref{fiteq} (dashed lines).  The resulting fitting parameters are given in Table.~\ref{tab4}.  }
\label{fit_density}
\vspace{0cm}
\end{figure}

\begin{figure}
\center
\includegraphics[scale=0.7]{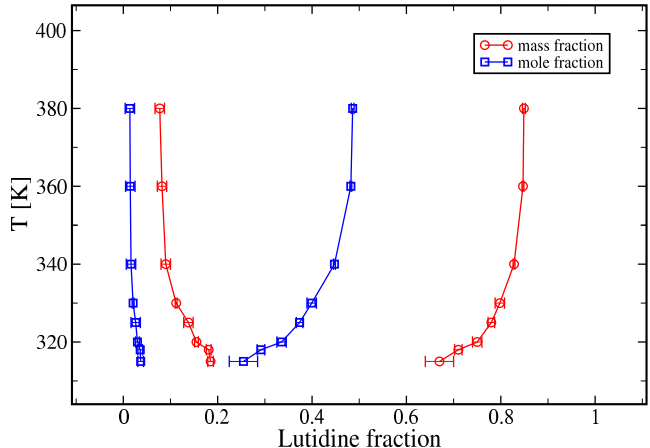}
\caption{{\color{black}Temperature versus} 2,6-lutidine mass fractions  (circles) given in  Table.~\ref{tab4},  and mole  fractions (squares) {\color{black}for the poor and rich phases  } obtained using Eq.~\ref{convert}. }
\label{phasediagram}
\end{figure}

To quantify the phase separation, the mass fraction of 2,6-lutidine $w_{lut}(z)$ has been calculated as a function of  $z$-coordinate from simulations at different temperatures, see Fig.~\ref{fit_density}.  
{\color{black} 'Classical` theories for the interface separating two coexisting phases such  Cahn and Hilliard \cite{CaH} or Landau-Ginzburg theory, predict a hyperbolic-tangent shaped  interfacial density profiles. This} has later been verified in simulations  of interfaces ~\cite{Inglesfield:95:00,chepela:00:00,muller:95:00,mecke:97:00,
watanabe:12:00,Omelyan:05:00,Blas:08:00,Yang:05:00,Eckelsbach:15:00}. 
{\color{black}  Since there are two interfaces in the present system, we  }
 fit the density profile to the function 
\begin{eqnarray}
w_{lut}(z)= w_{lut}^p+\frac{w_{lut}^r-w_{lut}^p}{2} \bigg [  \tanh\bigg(\frac{z-z_0+c}{\lambda}\bigg)- \tanh \bigg(\frac{z-z_0-c}{\lambda} \bigg) \bigg ],
\label{fiteq}
\end{eqnarray}
with $w_{lut}^r$ and $w_{lut}^p$ being
the mass fractions  of 2,6-lutidine in the lutidine-rich and lutidine-poor  phases. $\lambda$ is a measure of the width of the interface (softness of the transition between the two regions) 
and is  proportional to the correlation length $\xi$, which is defined from the decay of the density-density correlation function. {\color{black}$c $ is half width of the lutidine-rich region} and $z_0$ is center of the lutidine-rich phase.
 Fits Eq.~\ref{fiteq} to the  profiles are shown in Fig.~\ref{fit_density} as  dashed lines. The parameters obtained from the fits are given in  Table.~\ref{tab4} while the coexistence curve {\color{black} obtained from} the {\color{black} data in the} table is shown in Fig.~\ref{phasediagram}.
In order to compare with  experiments~\cite{andon:52:00,Grattoni:93:00,cox:56:00} one may need to convert  between mass fractions and  mole fractions. The appropriate equations {\color{black} for this} are
\begin{eqnarray}
x_{i}=\frac{w_{i}}{w_{i}+(1-w_{i})\frac{M_{i}}{M_{j}}}, \hspace{2cm} w_{i}=\frac{x_{i}}{x_{i}+(1-x_{i})\frac{M_{j}}{M_{i}}},
\label{convert}
\end{eqnarray}
where $w,x$ indicates mass and mole fractions respectively, while $i,j$ refers to water and 2,6-lutidine molecules with molar masses $M_{i,j}$.  
The mass and mole fractions of 2,6-lutidine are shown in the left and right vertical axes in Fig.~\ref{fit_density} for several temperatures, and by circles and squares  in Fig.~\ref{phasediagram}.

\begin{table}
\centering
\vspace{1cm}
\begin{tabular}{|l| c|c|c|c|c|c|c|}
\hline
 T[K]  & $w_{lut}^r  $ lutidine-rich phase   & $w_{lut}^p $ lutidine-poor phase     & $\lambda$ [nm]  &c [nm]  \\
         \hline
315 & 0.67 $\pm$ 0.03 &   0.185 $\pm$ 0.006 &   3. $\pm$ 0.8   & 3.4 $\pm$ 0.2 \\
      \hline
318 & 0.71 $\pm$ 0.008 &   0.181 $\pm$ 0.005 &   1.82 $\pm$ 0.2  & 3.75 $\pm$ 0.02 \\
      \hline    
320 & 0.75  $\pm$ 0.01 &   0.155 $\pm$ 0.004&  1.8 $\pm$ 0.3  & 4.20 $\pm$ 0.02 \\
   \hline
325 & 0.78  $\pm$ 0.007 &   0.138 $\pm$ 0.01&   1.6  $\pm$ 0.2  & 4.39 $\pm$ 0.03 \\
   \hline   
330 & 0.798 $\pm$ 0.01&  0.112  $\pm$ 0.002 & 1.38 $\pm$ 0.2  &  5.05 $\pm$ 0.01 \\
\hline
340 &0.828 $\pm$0.002 &  0.090 $\pm$ 0.01 &  1.18  $\pm$ 0.1  & 5.25 $\pm$ 0.02 \\
\hline
360 &   0.847 $\pm$ 0.002  &    0.082 $\pm$ 0.01 & 1.08  $\pm$ 0.02 & 5.49 $\pm$ 0.001\\
\hline
380 & 0.849 $\pm$ 0.003 &  0.077  $\pm$ 0.01& 1.05  $\pm$0.05 &   5.58 $\pm$ 0.02\\
\hline
\end{tabular}
\caption{Lutidine mass fraction   and the correlation length   from fits the simulation results to Eq.~\ref{fiteq}  versus temperature.}
\label{tab4}
\end{table}

{\color{black} 
When $T_c$ is approached from above, $\lambda$ increases and will eventually not be much smaller than $c$.
Then, the system will be too small to accommodate a saturated  lutidine-rich phase. We note that
this starts to occur at $T \approx 320$~K. This makes it difficult to determine the lutidine mass fraction $w^r_{lut}$  in the lutidine-rich  phase. At higher temperatures we could just read off  the value of  $w^r_{lut}$ from the flat part of the mass fraction profiles.
The alternative way to determine $w^r_{lut}$  is    by fitting the parameters in Eq.~\ref{fiteq} to simulation data. At high temperatures, this method   gives the same values for $w^r_{lut}$   as  the ones extracted from the flat parts of the mass fraction profiles, whereas  for $315$ and $318$~K it gives the values that are higher than the maxima in Fig.~\ref{fit_density}. The question is now whether   this procedure could be trusted or not. First we note  that the fit of the mass fraction profile to the  functional  form of Eq.\ref{fiteq}   looks very good. We tested the validity of this procedure further
by running simulations of a smaller system in which the width of the lutidine-rich region is about one third of the present one, at two temperatures $330$ and $380$~K. For the smaller system the density profiles do not exhibit flat regions corresponding to the  lutidine-rich phases at these temperatures, unlike the original simulations. By fitting  the mass fraction profiles to Eq.~\ref{fiteq}, we do however obtain  
 the same values for  $w^r_{lut}$   in the lutidine-rich regions as in the large system at the same temperatures,  despite that  the maximum of the mass fraction  profiles are about $20\%$  smaller  compared to the larger system.
 Although we do know that it eventually may break down close to $T_c$, but we are clearly not close enough for that.} 
\\

{\color{black} Now, we turn to the critical properties of the system. We fully realize that an
accurate calculation of critical exponents and $T_c$ would require simulations
closer to $T_c$, larger systems and a finite-size scaling analysis \citep{barber:83:00}.
This would for the present fairly complicated system call for an unjustifiable
amount of computer resources. Therefore, based on our simulations, the  calculated 
exponents are obtained within certain amount of statistical errors.\\
 As mentioned earlier, in order to minimize the finite-size effects
we  limited our simulations to the range of temperatures further away from $T_c$, for which the bulk correlation length of the mixture is distinctively  smaller 
than the linear dimension of the simulation box (see below) and   the order parameter is relatively  large. Under such conditions we do not expect  the mean field behaviour  
to occur \cite{mon:94:00,pan}.  On the other hand,  in this temperature range the corrections  to the critical scaling are relevant and therefore we will employ them.
The critical point and the shape of the coexistence curve  were determined by using well established procedures \cite{guissani:93:00,vnuk:83:00,beysens:79:00,aharony:80:00,bagnuls:81:00,durovn:04:00,vega:06:00}. } 
By employing the Wegner expansion~\cite{wegner:72:00}, the order parameter (OP) which in this case is  the mass fraction difference of lutidine in the rich and poor phases $w_{lut}^r-w_{lut}^p$, can be written as 
\begin{eqnarray}
w_{lut}^r-w_{lut}^p=B_0 \tau ^\beta+B_1 \tau^{\beta+\Delta},
\label{OP}
\end{eqnarray}
 where $\tau= \frac{T-T_c}{T_c}${\color{black}. The } rectilinear diameter can be similarly expressed  as
\begin{eqnarray}
\frac{w_{lut}^r+w_{lut}^p}{2}=w_{c}+ A_0 \tau +A_1 \tau^{1-\Gamma}.  
\label{rect}
\end{eqnarray}

\begin{figure}
\center
\includegraphics[scale=0.62]{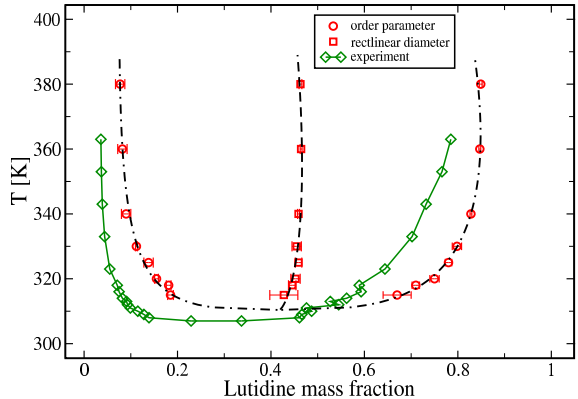}
\caption{ The coexistence  curve (dash-dotted line) determined by fitting Eq.~\ref{OP} to the order parameter data (circles) from  simulations,  and  the rectilinear diameter
obtained  by fitting  Eq.~\ref{rect} (dashed line) to the simulation data (squares).
The parameters of the fit are given in the first row of Table.~\ref{rect_fit_2}. The green diamond symbols represent the experimental data~\cite{stephenson:93:00}.     }
\vspace{1cm}
\label{coex}
\end{figure}

\begin{figure}
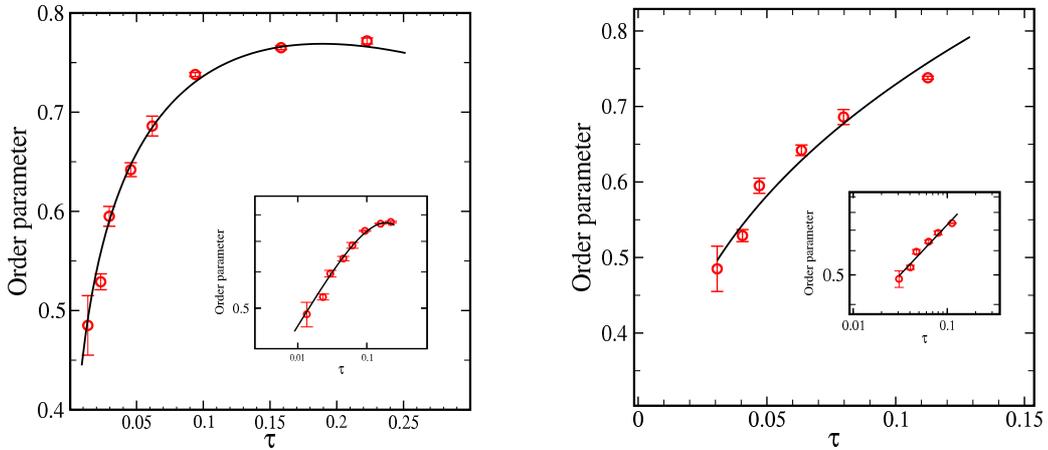

\center
\vspace{2cm}
\includegraphics[scale=0.75]{OP.eps}
\hspace{1cm}
\includegraphics[scale=0.75]{Fit_firstterm_Phasediagram.eps}
\caption{{\color{black} (left) The order parameter obtained from simulations (symbols) versus $\tau$ together with a fit to Eq.~\ref{OP} (parameters from the first row of  Table.~\ref{rect_fit_2}). The inset is in log-log scale. The curvature at the higher reduced temperatures in the inset shows the importance of  correction-to-scaling 
taken into account by exponent $\Delta$. (right) The order parameter  versus $\tau$, together with a fit to  Eq.~\ref{OP} without correction-to-scaling, $B_1=0$, and $\beta$ fixed to  the Ising value ($0.326$), while we ignored two temperatures  furthest away from $T_c$ (inset is in log-log scale).  } }
\vspace{1cm}
\label{op}
\end{figure}

 In {\color{black}Eqs.~\ref{OP}-\ref{rect} } $w_{c}$ is the mass fraction of lutidine at the critical point,  $A_{i}$,  $i=0,1$ and $B_{i}$, $i=0,1$ are {\color{black} non-universal } constants, while $\beta$, $\Delta$ and $\Gamma$ are universal critical exponents. For the $3D$ Ising universality class relevant  for the  present study, the  exponents are approximately $\beta=0.326$, $\Delta=0.5$ and
  $\Gamma=0.1$~\cite{Pelissetto:2002,gitterman,mattis,brankov}.
 We  fitted the obtained $w_{lut}^p$ and $w_{lut}^r$  from the simulations (Table.~\ref{tab4})   to  Eqs.~\ref{OP}-\ref{rect}. 
 This resulted in the values 
of $T_c$ and  $A_i$ and $B_i, i=0,1$  in  the first line of Table.~\ref{rect_fit_2}, when the three exponents were fixed to  their $3D$  Ising universality class values.
The results of the fit corresponding to the first row of the Table.~\ref{rect_fit_2} are shown in Fig.~\ref{coex} together with the experimental coexistence curve ~\cite{stephenson:93:00}.  The obtained curve exhibits a slight shift to the right as  compared to the experimental one.
{\color{black}In Fig.~\ref{op} we show the OP  as a function of the reduced temperature together with the fit to Eq.~\ref{OP} with (left panel) and without (right panel) correction-to-scaling term.
One can see that the OP is  fitted quite well to the power law with  the  3D Ising exponent $\beta$ (after ignoring two data points furthest away from 
$T_c$); including the correction-to scaling  makes this fitting  work also for temperatures further
away from $T_c$.
 As a check for consistency of our estimates, we  performed fitting treating the exponent $\beta$ as a free parameter.
This  fit provides a value of $\beta$ that is only slightly different from the $3D$ Ising exponent (see Table.~\ref{rect_fit_2}).

We also estimated the  temperature interval in which the UCP of the mixture is located. This was
 done by running simulations  (with  the  similar setting as  in the original ones) for   a smaller system  ($L_x= 5nm ,L_y=5nm ,L_z=12 nm$) at several higher temperatures.  The
simulations data show a phase-separated mixture at $450$~K, while a  mixed liquid phase at
$510 K$. This indicates that UCP is located between $450$-$510$~K, in agreement with experiments ~\cite{andon:52:00,cox:56:00,francis:61:00}.
 }
 
\begin{table}
\center
\vspace{1cm}
\scalebox{1}{
\begin{tabular}{|c  |c  |c  |c | c  |c  |c | c | c |}
\hline
   $\beta $ &   $\Gamma$ & $ \Delta$ & $T_c~$[K]   &    $ w_{c}$ &   $A_0$    &    $A_1$  &    $B_0$  & $B_1$ \\
      \hline
0.326 (fix)&    0.1 (fix) &   0.5 (fix)&     310.8 $\pm$ 0.5&    0.42   $\pm$ 0.002&   -2.84  $\pm$ 0.1 &    2.6  $\pm$ 0.1 &   2.18  $\pm$ 0.05 &     -1.95  $\pm$ 0.1  \\
\hline
0.34 $\pm$ 0.06 &    0.1 (fix)  &   0.5 (fix)&    310.5 $\pm$ 1.5 &    0.41 $\pm$ 0.01 &   -2.9 $\pm$ 0.2&   2.7  $\pm$ 0.2 &  2.3 $\pm$ 0.5 &     -2.2 $\pm$ 0.7\\  
\hline
\end{tabular}}
\caption{{\color{black} Parameters obtained when fitting Eqs.~\ref{OP} and \ref{rect} to the mass fractions $w_{lut}$ resulted from simulations.  In the first
row $\beta$, $\Gamma$ and $\Delta$ were fixed to the $3D$  Ising values, while  $\beta$ is treated as a free parameters in the second row.}
 }
\label{rect_fit_2}
\end{table}

\subsection{The surface tension  and the correlation length  of  2,6-lutidine/water mixture}

\begin{figure}
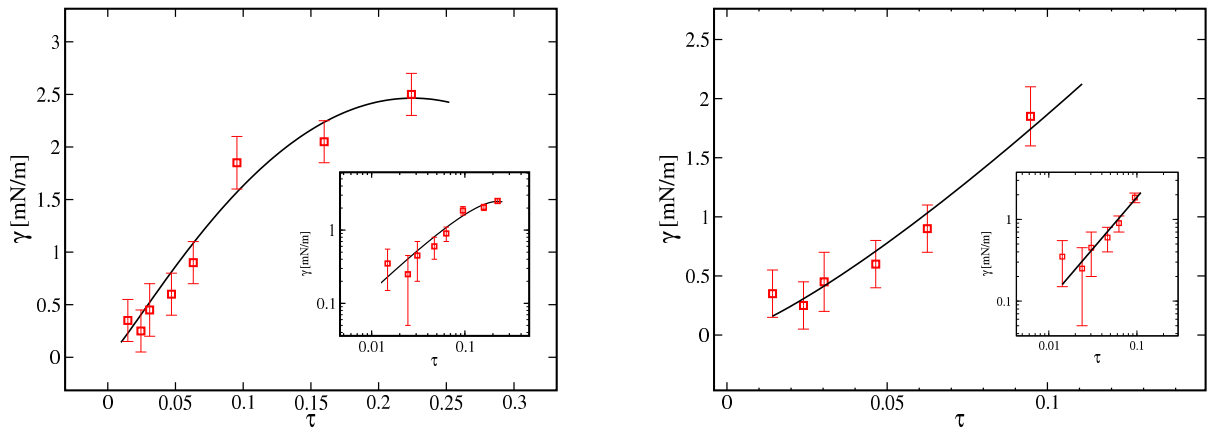

\center
\vspace{1cm}
\includegraphics[scale=0.67]{ST.eps}
\hspace{1cm}
\includegraphics[scale=0.67]{ST_firstterm.eps}
\caption{{\color{black} (left) The surface tension  of 2,6-lutidine/water  extracted from  simulations (symbols) together with a fit  to  Eq.~\ref{STexp} (parameters from the first row in Table.~\ref{STtab}). The inset is in log-log scale. 
(right) The surface tension versus $ \tau$ determined by a fit to Eq.~\ref{STexp} without correction-to-scaling, $C_1=0$, and ignoring  two temperatures  furthest away from  $T_c$ . } }
\label{ST}
\end{figure}

\begin{figure}
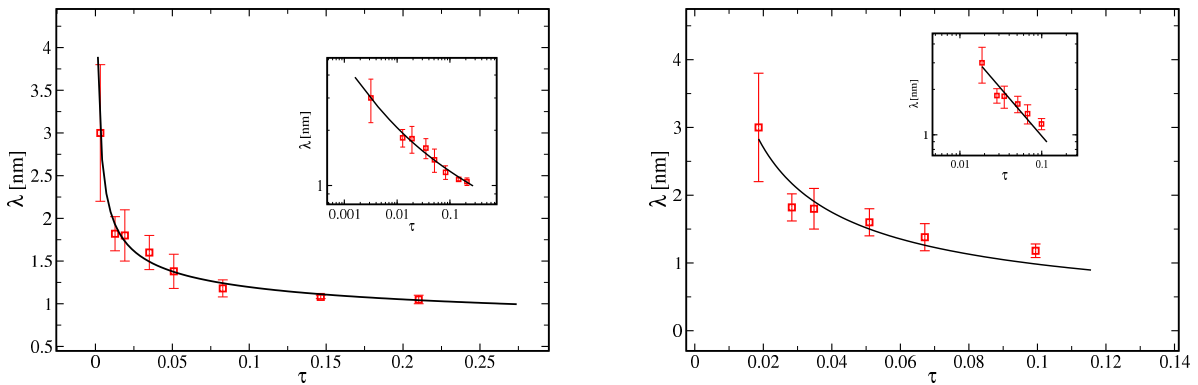

\center
\vspace{1cm}
\includegraphics[scale=0.6]{Correlationlenght.eps}
\hspace{1cm}
\includegraphics[scale=0.6]{Correlationlenght_firstterm.eps}
\caption{{\color{black} (left) The thickness of the interface, $\lambda$,  extracted from  simulations (symbols) together with a fit  to  Eq.~\ref{correxp} (parameters from the first row in Table.~\ref{Correlationlenght_tab}). The inset is in log-log scale. (right) The thickness of the interface versus $ \tau$ determined by a fit  to  Eq.~\ref{correxp} without correction-to-scaling, $\lambda_1=0$, and ignoring two temperatures  furthest away from  $T_c$ .}}
\label{Correlationlenght}
\end{figure}
We computed  the surface tension in the phase-separated system (above the LCT)   as a function of temperature from the simulations {\color{black} as an} integral of the difference between the normal and tangential  component{\color{black}s} of the  pressure (stress) tensor across the interface \cite{rowlinson:82:00}; the result as a function  the reduced temperature is plotted in Fig.~\ref{ST}. The surface tension and the  correlation length have 
similar scaling behaviour as the OP, but with different universal exponents~\cite{rathjen:77:00,aharony:80:00,beysens:82:00,rowlinson:82:00}
 \begin{eqnarray}
\gamma=C_0  \tau^{\mu} +C_1 \tau^{\mu+\Delta},
\label{STexp}
\end{eqnarray}
 \begin{eqnarray}
\lambda= \lambda_0 \tau^{-\nu} +\lambda_1 \tau^{-\nu+\Delta} ,
\label{correxp}
\end{eqnarray}
where $C_i$, $i=0,1$ and $ \lambda_i$, $i=0,1$ are non-universal constants.  
{\color{black} We fitted the simulation data of $\gamma$ and $\lambda$ to these expressions 
using the value of $\Delta$   fixed to its $3D$  Ising universality class value.}
In the Tables.~\ref{STtab}-\ref{Correlationlenght_tab}, the first rows are fits with {\color{black}the}  exponents $\mu$ and $\nu$ {\color{black}fixed} to the $3D$  Ising universality class values, {\color{black}while} the second rows show the  fits with $\mu$ and $\nu$ as free fitting parameters{\color{black}. The} last rows {\color{black}show fits  obtained} by imposed $T_c= 310.5 \pm 1.5$ as estimated from the coexistence curve fit (Table.~\ref{rect_fit_2}).
{\color{black}All these different way of fitting give similar values of parameters.} 
The fits to Eqs.~\ref{STexp}-\ref{correxp} with parameters from first rows of  Tables.~\ref{STtab}-\ref{Correlationlenght_tab} (fixed critical exponents $\mu$ and $\nu$) 
are shown in  Figs.~\ref{ST}-\ref{Correlationlenght}.\\

\begin{table}
\centering
\vspace{1cm}
\begin{tabular}{|c| c |c| c |c |  }
\hline
   $\mu $ & \hspace{0.2cm}   $\Delta$ & \hspace{0.2cm}   $C_0 $[mN/m]    & \hspace{0.2cm}   $C_1 $ [mN/m]  &  \hspace{0.2cm}  $T_c$ [K]   \\
      \hline
1.26 (fix)&\hspace{0.2cm}   0.5 (fix)  &\hspace{0.2cm}     56.8  $\pm$3 &\hspace{0.2cm}   -85.7  $\pm$ 7 &  \hspace{0.2cm} 311.45  $\pm$ 0.75  \\
\hline
1.28 $\pm$ 0.18   &\hspace{0.2cm}   0.5  (fix) &\hspace{0.2cm}    65  $\pm$ 23& \hspace{0.2cm}   -100 $\pm$ 40 &  \hspace{0.2cm} 311.5 $\pm $ 2.5   \\ 
\hline
1.35 $\pm$ 0.11   &\hspace{0.2cm}   0.5 (fix)  &\hspace{0.2cm}    73  $\pm$ 16& \hspace{0.2cm}   -113 $\pm$ 27 &  \hspace{0.2cm} 310.5 $\pm $ 1.5  (fix) \\
\hline
\end{tabular}
\caption{{\color{black} Parameters obtained when fitting Eq.~\ref{STexp} to the surface tension resulted from simulations.  In the first row, the critical exponents $\mu$ and $\Delta$ are fixed, while the second row shows fit with $\mu$ being a free parameter.  In the last row the fit has been done by restricting $T_c$ to the interval $[309-312]$~K as estimated from coexistence curve fits (Table.~\ref{rect_fit_2}).} }
\label{STtab}
\end{table}

\begin{table}
\centering
\vspace{1cm}
\begin{tabular}{|c|  c | c|  c |  c | c|  c| }
\hline
   $\nu $ &  \hspace{0.2cm}  $\Delta$ &  \hspace{0.2cm}  $\lambda_0$ [nm]      &  \hspace{0.2cm}  $\lambda_1$ [nm]     &  \hspace{0.2cm}  $T_c$ [K]   \\ 
      \hline
0.63 (fix)& \hspace{0.2cm}   0.5 (fix)& \hspace{0.2cm} 0.036  $\pm$ 0.013 &   \hspace{0.2cm}  0.77  $\pm$0.05 &  \hspace{0.2cm} 314    $\pm$ 0.4  \\ 
      \hline
0.62 $\pm$ 0.02  &\hspace{0.2cm}   0.5  (fix) &\hspace{0.2cm}    0.06   $\pm$ 0.04& \hspace{0.2cm}   0.67 $\pm$ 0.09 &   \hspace{0.2cm} 313.3 $\pm $ 0.8      \\
\hline
0.68 $\pm$ 0.05  &\hspace{0.2cm}   0.5 (fix) &\hspace{0.2cm}    0.087   $\pm$ 0.003& \hspace{0.2cm}   0.48 $\pm$ 0.07 &   \hspace{0.2cm} 310.5 $\pm $ 1.5  (fix)    \\
\hline
\end{tabular}
\caption{{\color{black} Parameters obtained when fitting Eq.~\ref{correxp} to the thickness of the interface resulted from  simulations. In the first row, the critical exponents $\nu$ and $\Delta$ are imposed to the fits,  while in the second and third rows the fits were done with $\nu$ as a free  parameter.  In the last row the fit has been done  by restricting $T_c$ to the interval  $[309-312]$~K as estimated from coexistence curve fits (Table.~\ref{rect_fit_2}).}   }
\label{Correlationlenght_tab}
\end{table}
       
\begin{table}[h]
\begin{center}
\begin{tabular}{| l | c  |c | c  |c |}
\hline
\hskip1cm Exponent & Simulation &\hskip0.3cm   $3D$ Ising & \hskip0.1cm $2D$ Ising & \hskip0.1cm Mean-field \\
\hline                                                                                                               
Order parameter, $\beta$ & $0.34\pm0.06$ &  $\sim$ 0.326&0.125& 0.5 \\
Correlation length, $\nu$  &$0.62\pm0.02$& $\sim$ 0.63& 1&0.5   \\ 
Surface tension, $\mu$ &$1.28\pm0.18$& $\sim$ 1.26& 1&1.5\\
\hline
\end{tabular}
\end{center}
\caption{The critical exponents obtained from the simulations compared to those of the 2- and 3-dimensional Ising model and mean field theory. }
\label{tab:exponents}
\end{table}

\subsection{Interactions between water and 2,6-lutidine in the mixture}

\begin{figure}
\center
\vspace{1cm}
\includegraphics[scale=0.525]{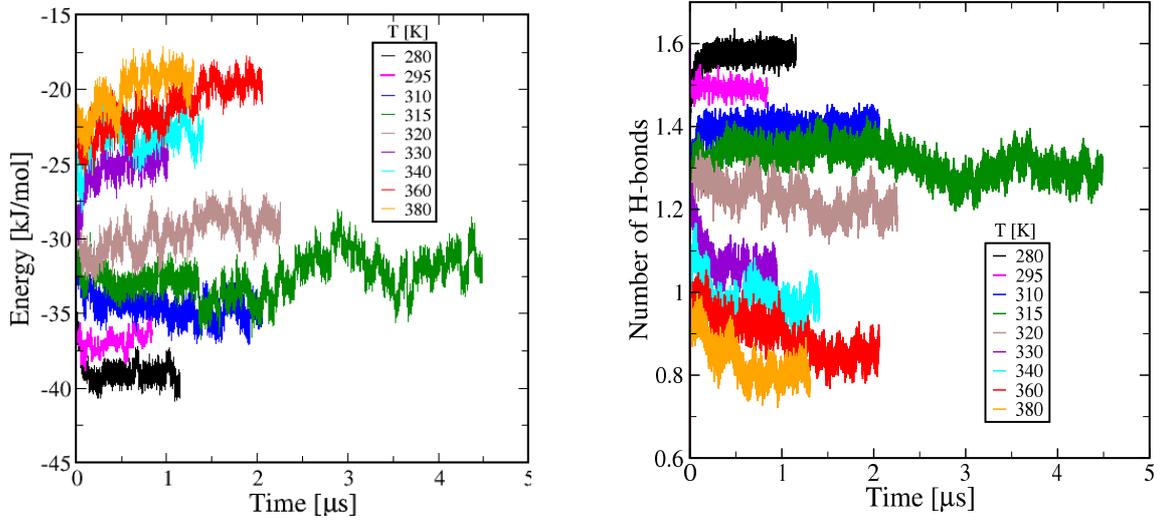}
\hspace{1cm}
\includegraphics[scale=0.8]{Hbond_1Dec.eps}
\caption{The energy (left) and the number of hydrogen bonds (right) between water and 2,6-lutidine molecules per lutidine molecules versus time.}
\vspace{1cm}
\label{LJ+Coul}
\end{figure}

Fig.~\ref{LJ+Coul} (left)  shows the interaction energy between water and 2,6-lutidine molecules per lutidine molecule for different temperatures. The figure indicates that the attraction between water and lutidine becomes stronger upon decreasing temperature.   Fig.~\ref{LJ+Coul} (right) shows the number of hydrogen bonds between water and 2,6-lutidine molecules per lutidine molecule. From the figure it is seen  that upon decreasing temperature  the number of hydrogen bonds between water and  2,6-lutidine molecules increases, in line with the behaviour seen in Fig.~\ref{LJ+Coul} (left).

\section{Conclusion}
 {\color{black}
  In this study, we have proposed an atomistic description of the 2,6-lutidine molecule which  we have shown  is able to successfully describe bulk 2,6-lutidine liquid. We then have employed this
model together with the TIP4P/2005 water model to study the phase behaviour of 
2,6-lutidine/water mixture near the LCP. We conclude that  by using  these models for molecules it is possible to  describe  the critical properties of the mixture well.  From the density profiles computed  in simulations we have obtained phase diagram of the mixture with the lower critical temperature $310.5\pm 1.5 K$, which is
just a couple of degrees higher than the experimental value ~\cite{andon:52:00,Grattoni:93:00,cox:56:00,francis:61:00}. We have found that the UCP is
located between $450$-$510$~K, in agreement with experiments ~\cite{andon:52:00,
cox:56:00,francis:61:00}. We have computed
the order parameter, the surface tension and the correlation length as a function of temperature. As expected, the order parameter and the surface tension
vanish upon approaching the LCP from above, while the correlation length increases.
Moreover, we have found  that close to  $T_c$ the temperature dependence of these quantities is well described by power laws. The calculated exponents deviate less than about $0.02$ from those
of the 3D Ising universality class~\cite{Pelissetto:2002,gitterman,mattis,brankov} to which the studied system belongs. However,
the estimated errors $[0.02-0.18]$ are clearly larger than this. A more accurate calculation
of the critical exponents and of Tc would require simulations closer to Tc , larger systems
and a finite-size scaling analysis \cite{barber:83:00}.
 
}

\section{Acknowledgments}
\label{Acknowledgement}
The work has been supported by the Swedish National Infrastructure for Computing (SNIC) with computer timed for the Center for High Performance Computing (PDC) and High Performance Computing Center North (HPC2N) and by National Science Center (Harmonia Grant No. 2015/18/M/ST3/00403). FP and OE would like  to thank J. Lidmar  and M. Wallin  for useful discussions. FP  would like to acknowledge E. Lindahl and B. Hess and their group members for helpful discussions during the meetings in SciLifeLab, and from L. Lundberg for quantum chemistry simulations of lutidine molecule.


\begin{thebibliography}{99}
%
\bibitem{Hirschfelder:1937} J. Hirschfelder, D. Stevenson, and H. Eyring, J. Chem. Phys. {\bf 5}, 896 (1937).
%
\bibitem{Narayanan:1994} T. Narayanan and A. Kumar,  Physics Reports {\bf 249}(3), 135 (1994).
%
\bibitem{Walker:1980} J. S. Walker and C. A. Vause, Phys. Lett. A {\bf 79}, 421 (1980).
%
\bibitem{Walker:1987} J. S. Walker and C. A. Vause, Sci. Am. {\bf 256}(5), 98-105 (1987).
%
\bibitem{Walker:1983} J. S. Walker and C. A. Vause,  J. Chem. Phys. {\bf 79}, 2660 (1983).
%
\bibitem{Almarza:2012} N. G. Almarza, Phys. Rev. E {\bf 86}, 030101 (2012).
%
\bibitem{Brovchenko:1997} I. V. Brovchenko and A. V. Oleinikova, J. Chem. Phys. {\bf 106}, 7756 (1997).
%
\bibitem{Robertson:2015} A. E. Robertson, D. H. Phan, J. E. Macaluso, V. N. Kuryakov, E. V. Jouravleva, C. E. Bertrand, I. K. Yudin, and M. A. Anisimov, Fluid Phase Equilibria {\bf 407}, 243 (2016).
%
\bibitem{Brovchenko:2001} I. V. Brovchenko and B. Guillot, Fluid Phase Equilibria {\bf 183-184}, 311 (2001).
%
\bibitem{MDS:2016} M. D. Smith, B. Mostofian, L. Petridis, X Cheng and J. C. Smith, J. Phys. Chem. B. {\bf 120}, 740 (2016).
%
\bibitem{Pohl_et} D. W. Pohl and W. I. Goldburg, Phys. Rev. Lett. {\bf 48}, 1111 (1982); L. Siegl and W. Fenzel, {\it ibid} {\bf 57}, 2191 (1986).
%
\bibitem{Frisken:1991} B. J. Frisken, F. Ferri, and D. S. Cannel, Phys. Rev. Lett. {\bf 66}, 2754 (1991). 
%
\bibitem{Broide:1993} M. L. Broide, Y. Garrabos, and D. Beysens, J. Chem. Phys. {\bf 47}, 3768 (1993).
%
\bibitem{Gallagher:1992} P. D. Gallagher and J. V. Maher, Phys. Rev. A {\bf 46}, 2012 (1992).
%
\bibitem{Beysens:1985} D. Beysens and D. Est\`ve,  Phys. Rev. Lett. {\bf 54}, 2123 (1985); D. Beysens and T. Narayanan, J. Stat. Phys. {\bf 96}, 997 (1999).
%
\bibitem{Kurnaz:1995} M. L. Kurnaz and J. V. Maher, Phys. Rev. E {\bf 51}, 5916 (1995).
%
\bibitem{Kurnaz:1997} M. L. Kurnaz and J. V. Maher,  Phys. Rev. E {\bf 55}, 572 (1997).
%
%
\bibitem{Hertlein-et:2008}
C. Hertlein, L. Helden, A. Gambassi, S. Dietrich, and C. Bechinger, Nature {\bf 451}, 172 (2008).
%
\bibitem{Gambassi-et:2009}
A. Gambassi, A. Macio\l{}ek, C. Hertlein, U. Nellen, L. Helden, C. Bechinger,  and S. Dietrich, Phys. Rev. E {\bf 80}, 061143 (2009).
%
\bibitem{Nellen-et:2009} U. Nellen, L. Helden, C. Bechinger, EPL {\bf 88}, 26001 (2009).
%
\bibitem{Nguyen-et:2013} V. D. Nguyen, S. Faber, Z. Hu, G. H.  Wegdam, and P. Schall, Nat. Commun. {\bf 4}, 1584 (2013).
%
\bibitem{andon:52:00} R. J. L. Andon and    J. D. Cox, J. Chem. Soc., 4601-4606 (1952).
%
\bibitem{stephenson:93:00} R. M. Stephenson, J. Chem. Eng. Data {\bf 38}(3), 428-431 (1993).
%
\bibitem{Grattoni:93:00} C. A. Grattoni, R. A. Dawe, C. Yen Seah, and J. D. Gray, J. Chem. Eng. Data. {\bf 38}, 516-519 (1993).
%
\bibitem{cox:56:00} J. D. Cox and E. F. G. Herington, Trans. Faraday Soc. {\bf 52}, 926-930 (1956).
%
\bibitem{Jayalakshmi:1993} Y. Jayalakshmi, J. S. V. Duijneveldt, and D. Beysens, J. Chem. Phys. {\bf 100}, 604 (1993).
%
\bibitem{francis:61:00} A. W. Francis, {\it {Critical Solution Temperatures}}, Advances in
Chemistry Series 31; American Chemical Society: Washington DC, (1961).
%
\bibitem{Loven:1963} A. W. Loven and  O. K. Rice, Trans. Faraday Soc. {\bf 59}, 2723 (1963). 
%
\bibitem{Guelari:1972} E. G\"ulari, A. F. Collings, R. L. Schmidt, and C. J. Pings, J. Chem. Phys. {\bf 56}, 6169 (1972). 
%
\bibitem{Levchenko:1993} V. A. Levchenko and V. P. Voronov, Int. J. Thermophys.  {\bf 14}, 221 (1993).
%
\bibitem{Mirzaev:2006} S. Z. Mirzaev, R. Behrends, T. Heinburg, J. Haller, and U. Kaatze, J. Chem. Phys. {\bf 124}, 144517 (2006). 
%
\bibitem{sadakane:11:00} K. Sadakane,  N. Iguchi, M. Nagao, H. Endo, Y. B. Melnichenko and H. Seto, Soft Matter {\bf 7}(16), 1334 (2011). 
%
\bibitem{sadakane:09:00} K. Sadakane, A.  Onuki, K. Nishida, S.  Koizumi and H. Seto, Phys. Rev. Lett. {\bf 103}, 167803 (2009). 
%
\bibitem{leys:13:00} J. Leys, D. Subramanian, E. Rodezno, B. Hammouda and
M. A. Anisimov, Soft Matter {\bf 9}, 9326 (2013). 
%
\bibitem{sadakane:14:00} K. Sadakane, H. Endo, K. Nishida and H. Seto, J. Solution Chem. {\bf 43}, 1722 (2014).
%
\bibitem{Onuki:04:00} A. Onuki and H. Kitamura, J. Chem. Phys.  {\bf 121}, 3143 (2004).
%
\bibitem{pousaneh:14:00} F. Pousaneh and A. Ciach, Soft Matter  {\bf 10}, 8188 (2014).
%
\bibitem{nellen:11:02} U. Nellen, J. Dietrich, L. Helden, S. Chodankar, K. Nygard, J. F. van der Veen and C. Bechinger, Soft Matter {\bf 7},  5360 (2011). 
%
\bibitem{bier:11:02} A. Gambassi, M. Oettel and S. Dietrich, Europhys. Lett. {\bf 95}, 60001 (2011). 
%
\bibitem{ciach:10:00} A. Ciach and A. Macio\l ek, Phys. Rev. E {\bf 81}, 041127 (2010).
%
\bibitem{pousaneh:12:00} F. Pousaneh, A. Ciach and A. Macio\l ek, Soft Matter {\bf 8}, 3567 (2012).
%
\bibitem{spoel:05:00} D. Van Der Spoel, E. Lindahl, B. Hess, G. Groenhof, A. E. Mark, and H. J. C.  Berendsen, J. Comput. Chem.  {\bf 26}(16), 1701 (2005). 
%
\bibitem{Schmid:11:00} N. Schmid, A. P. Eichenberger, A. Choutko, S. Riniker, M. Winger, A. E. Mark, and  W. F. van Gunsteren, Eur. Biophys. J. {\bf 40}, 843-856 (2011).
%
\bibitem{Darden:93:00} T. Darden, D. York, and L. Pedersen, J. Chem. Phys.  {\bf 98}(12), 10089 (1993).
%
\bibitem{Bussi:07:00} G. Bussi, D. Donadio, and M. Parrinello, J. Chem. Phys.  {\bf 126}, 014101 (2007).
%
\bibitem{Parrinello:81:00} M. Parrinello and A. Rahman, J. Appl. Phys.   {\bf 52}, 7182-7190 (1981).
%
\bibitem{Hess:97:00} B. Hess, H. Bekker,  H. J. C. Berendsen, and J. G. E. M. Fraaije, J. Comput. Chem.  {\bf 18}, 1463-1472 (1997).
%
\bibitem{Abascal:05:00} J. L. F Abascal and C. Vega, J. Chem. Phys.  {\bf 123}, 234505 (2005).
%



{\color{black} 
\bibitem{VMD} W. Humphrey, A.  Dalke and K.  Schulten,  J. Mol. Graphics  {\bf 14}, 33 (1996).

\bibitem{Giorgino} T. Giorgino, Comput.
Phys. Commun. {\bf 185},   317 (2014).


}



\bibitem{Steele:86:0} W. V. Steele, R. D. Chirico, A. Nguyen, and S. E. Knipmeyer, J. Chem. Thermodyn. {\bf 27}, 311-334 (1995).
%
\bibitem{bakshi:93:00} M. S. Bakshi, J. Chem. Soc., Faraday Trans. {\bf 89}, 3049  (1993).
%
\bibitem{Jenkins:90:00} J. O. Jenkins and    J. W. Smith, J. Chem. Soc. B,  1538-1541  (1990).
%
\bibitem{Althof:97:00} T. Mainzer-Althof  and D. Woermann, Ber. Bunsen-Ges. Phys. Chem.  {\bf  101 }, 1014  (1997).

\bibitem{NIST} V. K. Shen, D. W.  Siderius, and W. P. Krekelberg, NIST Standard Reference Simulation Website, NIST Standard Reference Database. 


\bibitem{feig:10:00} M. Feig, {\it {Modelling solvent environment: applications to simulations of biomolecules}}, Wiley-VCH, Weinheim, (2010).
%


\bibitem{berens:83:00} P. H. Berens, D. H. J. Mackay, G. M. White, and K. R. Wilson, J. Chem. Phys. {\bf 79}, 2375 (1983).




\bibitem{waheed:11:0} Q. Waheed and O. Edholm,  J. Chem. Theory Comput.  {\bf 7}, 2903-2909 (2011).


\bibitem{Allen:87:00} M. P. Allen and  D. J. Tildesly, {\it {Computer simulation of liquids}}, Oxford Science Publications, Oxford (1987).

\bibitem{Haynes} W. M. Haynes (ed.), {\it {CRC Handbook of Chemistry and Physics}}, 94th edition. CRC Press, Boca Raton,  (2013).


\bibitem{mortimer:08:00} R. G. Mortimer, {\it { Physical Chemistry} },  Elsevier Academic Press, San Diego, California, 3rd edition, (2008).






\bibitem{smith:65:00} N. O. Smith, J. Chem. Educ. {\bf 42}(12), 654 (1965).
{\color{black}
\bibitem{godavat} R. Godawat, S. N. Jamadagni, J. R. Errington  and S. Garde, Ind. Eng. Chem. Res.,  {\bf 47}(10), 3582  (2008).

\bibitem{shi:2011}Z. Shi, P. G. Debenedetti and F. H. Stillinger, J. Chem. Phys. {\bf 134}, 114524  (2011).


\bibitem{MacDowell:2006} L. G. MacDowell, V. K. Shen an J. R. Errington, J. Chem. Phys. {\bf 125} 034705 (2006).

\bibitem{chepela:00:00} G. A. Chapela, G. Saville, S. M. Thompson, and J. S. Rowlinson, J. Chem. Soc., Faraday Trans. 2,  {\bf 73}, 1133 (1977).


\bibitem{holcomb:00:00} C. D. Holcomb, P. Clancy, and J. A. Zollweg, Mol. Phys. {\bf 78}, 437 (1993).

\bibitem{chen:00:00} L. J. Chen, J. Chem. Phys. {\bf 103}, 10214 (1995).


\bibitem{trokhymchuk:00:00} A. Trokhymchuk and J. Alejandre, J. Chem. Phys. {\bf  111}(18), 8510 (1999).
{\color{black}
\bibitem{CaH} J. W. Cahn and J. E. Hilliard, J. Chem. Phys. {\bf 28}, 258 (1958).
}
\bibitem{muller:95:00} M. M\"uller, K. Binder, and W. Oed, J. Chem. Soc., Faraday Trans. {\bf 91}, 2369 (1995).


\bibitem{mecke:97:00} M. Mecke, J. Winkelmann, and J. Fischer, J. Chem. Phys. {\bf 107}, 9264 (1997).

\bibitem{watanabe:12:00} H. Watanabe, N. Ito, and C. Hu, J. Chem. Phys. {\bf 136}, 204102 (2012).

\bibitem{Omelyan:05:00} I. Omelyan, F. Hirata,  and A. Kovalenko, Phys. Chem. Chem. Phys. {\bf 7}, 4132-4137 (2005).
}

\bibitem{Blas:08:00} 
F. J. Blas, L. G. MacDowell, E. D. Miguel, and G. Jackson, J. Chem. Phys. {\bf 129}, 144703 (2008).

{\color{black}

\bibitem{Inglesfield:95:00} J. E. Inglesfield, {\it {Cohesion and Structure of Surfaces}}, Vol 4 edited by K. Binder, M. Bowker, J. E. Inglesfield and P. J. Rous, Elsevier Science, (1995).

}
\bibitem{Yang:05:00} 
T. H. Yang and C. Pan,  Intern. Journ. of Heat and Mass Transfer. {\bf 48}, 3516-3526 (2005).

{\color{black}
\bibitem{Eckelsbach:15:00} 
S. Eckelsbach and    J. Vrabec, Phys. Chem. Chem. Phys. {\bf 17}, 27195-27203 (2015).
}

\bibitem{barber:83:00} M. N. Barber in {\it Phase Transitions and Critical Phenomena},
Volume 8, edited by C. Domb and J. L. Lebowitz, Academic Press, New York, (1983), V. Privman, in {\it Finite Size Scaling and Numerical Simulation of
  Statistical Systems}, edited by V. Privman, World Scientific, Singapore, (1990).
 
{\color{black}
\bibitem{mon:94:00} K. Mon and  K. Binder, J. Chem. Phys. {\bf 96}, 6989 (1994).
\bibitem{pan} A. Z. Panaglotopoulos, International Journal of Thermophysics {\bf 15}, 1057 (1994).
}

\bibitem{guissani:93:00} Y. Guissani and B. Guillot, J. Chem. Phys. {\bf 98}(10), 8221 (1993).

{\color{black}
\bibitem{vnuk:83:00} F. Vnuk and J.  Chem. Soc., Faraday Trans. 2,  {\bf 79}, 41-55 (1983).

\bibitem{beysens:79:00} D. Beysens,  J. Chem. Phys.  {\bf 71}, 2557 (1979).
\bibitem{aharony:80:00} A. Aharony and G. Ahlers, Phys. Rev. Lett. {\bf 44}, 782 (1980).

\bibitem{bagnuls:81:00} C. Bagnuls and C. Bervillier,  Phys. Rev. B {\bf 24}, 1226-1235 (1981).
}
\bibitem{durovn:04:00} V. A. Durov, in: J. Samios, V.A. Durov (Eds.), {\it { Novel Approaches to the
Structure and Dynamics of Liquids: Experiments, Theories and Simulations} }, NATO Science Series. II, Mathematics, Physics and Chemistry, Kluwer,
Dortrecht, 17-40 (2004).



\bibitem{vega:06:00} C. Vega, J. L. F. Abascal, and I. Nezbeda, J. Chem. Phys. {\bf 125}, 034503 (2006).


\bibitem{wegner:72:00} F. Wegner, Phys. Rev. B {\bf 5}(11), 4529 (1972).


\bibitem{Pelissetto:2002}  A. Pelissetto and E. Vicari, Phys. Rep. {\bf 368}, 549 (2002).

\bibitem{gitterman} M.  Gitterman and V. Halpern, {\it {Phase Transition: A Brief Account with Modern
Applications}}, World Scientific, (2004).

\bibitem{mattis} D. C. Mattis, R. H. Swendsen, {\it {Statistical Mechanics Made Simple}}, 2nd ed. World Scientific,
(2008).


\bibitem{brankov} J. G. Brankov, D. M. Danchev, and N. S. Tonchev, {\it {Theory of Critical Phenomena in Finite-size Systems: Scaling and Quantum Effects}}, Singapore: World Scientific, (2010).


\bibitem{rowlinson:82:00} J. S. Rowlinson and B. Widom, {\it {Molecular Theory of Capillarity}}, Oxford University Press, (1982).







\bibitem{rathjen:77:00} W. Rathjen and J. Straub, {\it {Surface tension and refractive index of six refrigerants from triple point up to critical point}}, Proc. 7th Symposium on  Thermophysical  Properties, Washingron, USA, May 10-12, (1977).


\bibitem{beysens:82:00} D. Beysens, A. Bourgou, and P. Calmettes, J. Phys. Rev A. {\bf 26},  3589 (1982).





%
%







\end{thebibliography}
\end{document}